\documentclass[twocolumn,prl,amsmath,amssymb,showpacs]{revtex4}

\usepackage{bm}

\ifx\pdftexversion\undefined
  \usepackage[dvips]{graphicx}
\else
  \usepackage[pdftex]{graphicx}
\fi

\def \be{\begin{equation}}
\def \ee{\end{equation}}
\def \bmlett{\begin{mathletters}}
\def \emlett{\end{mathletters}}

\def \ve{\varepsilon}

\def \ra{\rightarrow}

\def \Z{\mathcal{Z}}
\def \J{\mathcal{J}}

\def \L{\mathcal{L}}

\begin{document}



\title{Back-action noise in strongly interacting systems: the dc SQUID and the interacting quantum point-contact}

\author{A. A. Clerk}
\affiliation{ Department of Physics, McGill University, Montr\'{e}al,
 Qu\'{e}bec, Canada, H3A 2T8}
\date{July 11, 2005}

\begin{abstract}

We study the back-action noise and measurement efficiency (i.e. noise temperature)
of a dc SQUID amplifier, and equivalently, a quantum point contact detector
formed in a Luttinger liquid.  Using a mapping to a dissipative tight-binding
model, we show that these systems are able to reach the quantum limit
even in regimes where several independent transport processes contribute
to the current.  We suggest how this is related to the underlying integrability
of these systems.

\end{abstract}
\maketitle

There has been considerable recent interest in studying detectors and amplifiers which add the minimal possible noise allowed by quantum mechanics to an input signal \cite{Gurvitz97, Levinson97, Aleiner97, Averin01, Korotkov01c, Pilgram02, Clerk03}.  Such detectors 
are necessary for single spin and gravity wave detection, as well as for applications related to quantum control and quantum computation.  In the important case where the detector is a mesoscopic conductor, understanding whether or
not one can achieve the quantum limit (i.e. have a minimal back-action effect) requires an understanding of both the output current noise of the system, as well as its back-action charge noise.  Theoretically, these quantities have been studied for a variety of
mesoscopic detectors, including non-interacting tunneling point-contact detectors \cite{Gurvitz97, Aleiner97} and more general coherent scattering detectors \cite{Pilgram02, Clerk03}.  
A general principle that has emerged from these studies is that reaching the quantum limit requires there to be no ``wasted information" in the detector-- there should be no other quantity besides the output variable of the detector that could be measured to reveal information on the input signal \cite{Clerk03}.

In this paper, we examine the ideality of mesoscopic detectors where non-trivial interactions are important.  An example system where this question is relevant is a quantum point contact detector (QPC) formed in an interacting Luttinger liquid.   QPC's are in widespread use as readouts of quantum-dot qubits; in the absence of interactions, they are known to be able to reach the quantum limit \cite{Gurvitz97, Korotkov01c, Pilgram02, Clerk03}.   What happens now when interparticle interactions are turned on?  Note that while the current noise of an interacting QPC has received considerable attention \cite{Kane94, Fendley95}, the corresponding back-action noise has not been studied.  

While an interacting QPC detector could be directly realized in a quantum Hall edge state, its study is also motivated by its connection to a second device of obvious practical importance, the dc SQUID amplifier.  While experimentally it is known that the dc SQUID can approach near quantum-limited operation, the theoretical limit for this system has
not been fully studied.  Previous studies either neglected the effect of the SQUID inductance \cite{Averin01}, or were based on the quantum Langevin equation \cite{Koch81,Danilov83}, an equation which is formally only valid in the limit of high temperatures or extreme dissipation \cite{Schmid82}.

In this paper, we calculate the back-action noise and measurement efficiency (i.e. noise temperature) of both the interacting QPC detector and the dc SQUID using controlled perturbative approaches.  We discuss how in each case, the principle of detection is the same:  the input signal modulates the tunneling of excitations across the detector.  In the QPC case, these excitations are electrons or quasiparticles.  In the SQUID case, the excitations are incoherently-tunneling Cooper pairs.   We address both the weak-tunneling (where leading order perturbation theory is valid) and strong-tunneling regimes of these systems.  In the latter regime a multitude of tunneling processes,
each transferring a different number of particles, will be significant.  As the output of the detector essentially averages over these processes, one would expect there to be lost information and hence excess back-action, much in the same way that multiple channels in a non-interacting QPC lead to a departure from the quantum limit \cite{Pilgram02, Clerk03}.  Surprisingly, we show that this is not the case.  Using a mapping to a dissipative tight binding model, we find that when one is in the scaling-limit of these systems, one remains at the quantum limit even when multiple tunnel processes contribute.  
We end by suggesting how this result is 
directly related to the integrability of the field theory describing these models.  

\begin{figure}
\center{\includegraphics[width=8 cm]{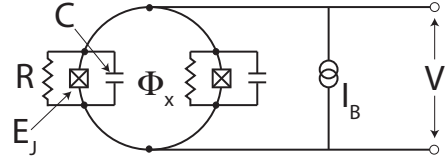}} \vspace{-0.5 cm}
\caption{\label{SQUIDPlot} Schematic of a dc SQUID amplifier.  Both
junctions are identical, and the SQUID loop has an inductance L.} \vspace{-0.5 cm}
\end{figure}

{\it Models}
We start by discussing the dc SQUID amplifier; Fig. 1 shows a schematic of a typical setup.  It consists of two identical Josephson junctions, each shunted by a resistance $R$ and capacitance $C$, placed in a ring of inductance $L$ which threads an external flux $\Phi_{x}$.  The junctions are symmetrically biased by a current bias $I_B$.  We consider the typical case where the SQUID is operated in a non-hysteretic mode, and where $I_B > 2 I_C$, where $I_C=2 e E_J / \hbar$ is the critical current of each individual junction.  The dc SQUID is a flux-to-voltage amplifier.  The input signal is a small additional flux $\Phi_{in}$ which adds to $\Phi_{x}$; by varying its value, one changes the voltage $V$ across the SQUID.  The back-action in this system is created by the circulating current $J$ around the SQUID loop \cite{Koch81}.   This current directly couples to the input flux, and thus its fluctuations act as a noisy back-action force on the system producing the input flux.

A convenient measure of the ideality of this system as a linear amplifier or detector is given by its noise temperature $k_B T_N$.  It quantifies the total noise added to the input by the amplifier, and includes the effect of back-action.  Quantum mechanics requires a certain minimal amount of back-action; as a result, there is a quantum limit on the noise temperature $k_B T_N \geq \hbar \omega /2$, where $\omega$ is the signal frequency \cite{Caves82}.  For an optimal coupling between the signal source and the SQUID, $T_N$ is determined by the noise properties of the uncoupled detector \cite{Clerk04c}:
\begin{equation}
	\chi \equiv
	\frac{k_B T_N}{\hbar \omega /2} = \sqrt{
		\frac{S_V(\omega) S_J(\omega) - S_{VJ}(\omega)^2}
			{(\hbar \lambda(\omega)/2)^2} }
		\geq 1
\end{equation}
Here, $S_V$ is the symmetrized noise in output of the detector (i.e. in $V$), $S_J$ is the symmetrized noise in the back-action force of the detector (i.e. $J$), and $S_{VJ}$ is the cross-correlation noise between these quantities.  $\lambda = d \langle V \rangle / d \Phi_{x}$ is the gain of the detector.  Note that for identical junctions $S_{VJ} = 0$ for the dc SQUID.  Also note that if our detector is used for quantum non-demolition qubit detection, $1/\chi^2$ represents the measurement efficiency ratio of the detector \cite{Clerk03}. 

We proceed to analyze the SQUID in a manner which highlights the role of Cooper pair tunneling.  Heuristically, the bias current $I_B$ will partially flow through the shunt resistors of the two junctions, and partially through the junctions.  The current through the junctions will be due to the incoherent tunneling of Cooper pairs \cite{Ingold99}:  Cooper pairs tunnel through the junction, simultaneously dissipating energy in the electromagnetic environment formed by the shunt impedances.  The voltage across the SQUID will then be set by the current flowing through the shunts: $V = (R/2) \left[I_B - (I_{CPT,1}+I_{CPT,2}) \right]$, where $I_{CPT,j}$, the Cooper-pair tunneling current through junction $j$ ($j=1,2$), depends both on the external flux and on the magnitude of the current bias.  In this picture, both the gain $\lambda$ and output noise $S_V$ can be directly related to $I_{CPT}$.  Similarly, the back-action circulating current will be simply given by the difference in the two Cooper-pair currents, $J = (I_{CPT,1} - I_{CPT,2})/2$.

The above picture can be made rigorous by using a standard Caldeira-Legget representation of the impedances in the SQUID, and writing down the path-integral representation of the Keldysh partition function $\Z$ for our system
\cite{Ingold99}.  Tracing out the environmental degrees of freedom, and removing the coupling to the bias current $I_B$ via a gauge transformation, $\Z$ may be written as 
$\Z = \int {\mathcal D} \theta {\mathcal D}  \phi  \exp \left(\frac{i}{\hbar} \int dt \L[\theta,\phi]\right)$. 
Letting $\phi = (\delta_1 + \delta_2)/2$ denote the average phase of the two junctions, and 
$\theta = (\delta_1 - \delta_2)/2 - \pi \Phi_x / \Phi_0$ describe their phase difference, we have:  
\begin{eqnarray}
	\frac{\L}{2} & = & 
		\sum_{\sigma = \pm}
		 \left[
			\sigma E_J \cos \left(\theta_\sigma(t) + \frac{\pi \Phi_x}{\Phi_0} \right)
				\cos \left(\phi_\sigma(t) + v t  \right)
			\right] 
	\nonumber \\
	&& 
		  + 
		 \sum_{\alpha = \phi,\theta}
		\int dt'	
			\left(
				\begin{array}{cc}
				  \alpha_c(t)   \\
				  \alpha_q(t)  \\
				\end{array}
			\right)
				 \hat{G}_\alpha^{-1}(t-t')
			\left(
				\begin{array}{cc}
				  \alpha_c(t')   \\
				  \alpha_q(t')  \\
				\end{array}
			\right)
\end{eqnarray}
The effective voltage bias $V_B = R I_B$ sets $v = 2 e V_B  / \hbar$, 
and we use the indices 
$c$ and $q$ to denote classical and quantum Keldysh fields \cite{Kamenev01}: $\phi_{\pm} = \phi_c \pm \phi_q$, $\theta_{\pm} = \theta_c \pm \theta_q$. The matrix Green functions $\hat{G}_{\alpha}$ ($\alpha = \theta, \phi$) are given in terms of the corresponding impedances $Z_{\alpha}$ via:
\begin{eqnarray}
	\frac{\hat{G}_{\alpha}(\omega)}{\hbar \omega} & = & 
	\left(
		\begin{array}{cc}
				-2 i  \textrm{Re} Z_{\alpha}(\omega) 
							\coth(\frac{\hbar \omega }{2 k_B T}) 
			&	- i  Z_{\alpha} (\omega) \\
				i  Z_{\alpha}^*(\omega) 	& 0 				\end{array}
	\right) 
\end{eqnarray}
with $Z_{\phi}(\omega) =   \left( 2/R + 2 i \omega C \right)^{-1}$ and  
$Z_{\theta}(\omega)  =   \left( 2/R + 2 i \omega C + 1/(i \omega L) \right)^{-1}$.   Note that $\textrm{Re} Z_{\theta}$ vanishes as $2 R (\omega L / R)^2$ for small R, implying that low-frequency fluctuations of $\theta$ will be suppressed.

Introducing sources to calculate $V$, $J$ and their fluctuations from the partition function, one can identify operators for the total Cooper pair current $I_{CPT}$ and the back-action current $J$.  As expected, they correspond, respectively, to the sum and difference of the Cooper-pair currents through each junction:
\begin{eqnarray}
	I_{CPT} & = & 2 I_C \cos(\theta + \pi \Phi_x / \Phi_0) \sin(\phi + v t) \\
	 I_{diff} & = &  I_C \sin(\theta + \pi \Phi_x / \Phi_0) \cos(\phi + v t)
\end{eqnarray}
From the path integral approach, one finds rigorously that 
$\langle V \rangle = (R/2)(I_B - \langle I_{CPT} \rangle)$; moreover, 
at zero temperature and zero frequency, one can directly relate the
noise correlators of interest to the Cooper-pair shot noise:  
$S_V = (R/2)^2 S_{I_{CPT}}$ and $S_J = S_{I_{diff}}$.  At finite frequency and or temperature, one finds additional contributions to the noise arising from the shunt impedances, as well as terms describing feedback between environmental fluctuations and Cooper-pair shot noise.  In what follows, we focus on the zero-temperature, zero-frequency limit; finite frequency and temperature effects will be discussed elsewhere.

At this point, we pause to point out the analogy between the dc SQUID and the interacting QPC detector.  The latter system is potential-to-current amplifier.  The signal of interest  modulates the tunneling amplitude $\Delta$ between two semi-infinite, spinless Luttinger liquid leads; the result is a corresponding modulation of the QPC current \cite{Kane92}.  In a bosonized representation, the tunnel Hamiltonian appears as a cosine potential $\mathcal{H}_t = -\Delta \cos \phi$,
and the Keldysh partition function for this system is identical to that for the dc SQUID {\it if} the phase $\theta$ is pinned.   To make this correspondence \cite{Weiss99}, we associate the dimensionless conductance 
$1/\rho = h / (2 e^2 R)$ with the Luttinger interaction parameter $g$, $2 E_J \cos \pi \Phi_x / \Phi_0$ with the QPC tunneling amplitude $\Delta$, and $2 V_B$ with the QPC bias voltage.  The current through the QPC will then correspond to $I_{CPT}$, and the back-action force corresponds to $I_{diff} / (I_C \sin \pi \Phi_x / \Phi_0)$.  
This correspondence also requires $Z_{\phi}$ be constant over frequencies of interest, implying
 that the cut-off frequency $\omega_C = 1 / (R C)$ be much larger than $v, e R I_C / \hbar$.  We can thus view the dc SQUID as an interacting QPC detector where, due to fluctuations in $\theta$, the magnitude of the tunnel matrix element is fluctuating.     

We now return to the SQUID, and address its back-action and measurement efficiency.  By expanding $\Z$ in powers of $E_J$, we can systematically describe processes involving multiple transfers of Cooper pairs.  Such an expansion is well controlled in the large-$I_B$ limit; for the experimentally relevant limit $\rho \ll 1$, the expansion converges for all $I_B >2 I_C$ \cite{Ingold99}.  The simplest limit is when only single Cooper-pair tunneling events play a role; for $\rho < 1$, this translates to the condition $I_C \ll I_B$.
In this regime, it is sufficient to calculate the noise and average voltage of the SQUID to lowest non-vanishing order in $E_J$ \cite{Averin01}.
Introducing the usual phase-phase correlation functions $\J(t)$:
\begin{eqnarray}
	\J_{\alpha}(t) = \frac{8 e^2}{h} \int_0^{\infty} \frac{d \omega}{\omega}
	\left( \textrm{Re }Z_{\alpha} (\omega) \right)  \left(e^{-i \omega t} -1\right)
\end{eqnarray}
we find that the measurement efficiency (or reduced noise temperature) for an optimal $\Phi_{x} = \Phi_0 / 4$ is given by:
\begin{eqnarray}
	\chi^2 = 
			\left(
		e^{2 \langle \theta^2 \rangle}
		\frac{
		\int_0^{\infty} dt \sin(v t) \textrm{Im}
			\left[
				 \exp\left(
					\J_{\phi}(t) + \J_{\theta}(t) \right)
			\right] 
	}{
		\int_0^{\infty} dt \sin(v t) \textrm{Im}
			\left[
				 \exp\left(
					\J_{\phi}(t) - \J_{\theta}(t) \right)
			\right] 
	} \right)^{2}
	\label{Eq:ChiFirstOrder}
\end{eqnarray}
with $ \langle \theta^2 \rangle = (8 e^2 / h) \int_0^{\infty} d\omega \textrm{Re} Z_{\theta}(\omega)/\omega$.
We see immediately that fluctuations in the phase $\theta$ prevent one from reaching the quantum limit (i.e. $\chi = 1$).  Heuristically, these fluctuations represent an extraneous source of noise:  the detector would function just fine even if $\theta$ was pinned.  The $\theta$ fluctuations both reduce the gain of the detector and increase the back-action noise.  Suppressing these fluctuations (i.e. pinning $\theta$) requires that the loop inductance $L$ be small.  More precisely, for $\rho \ll 1$, 
Eq. (\ref{Eq:ChiFirstOrder}) becomes:
 \begin{eqnarray}
	\chi^2 = \left(\frac{ Z_\phi(v) + Z_\theta(v)}
			   { Z_\phi(v) - Z_\theta(v)} \right)^2
\end{eqnarray}
Reaching the quantum limit in this large bias, weak tunneling regime thus requires that $Z_{\theta} / Z_{\phi}$ be small at the characteristic Josephson frequency $v$ set by the bias current.  Note that in the $L \ra 0$ limit, our system always reaches the quantum limit as long as tunneling is weak, irrespective of further details of the impedance $Z_\phi$.  As the SQUID is equivalent to a interacting QPC in this limit, we can also conclude that {\it an interacting QPC in the weak-tunneling regime always reaches the quantum limit, irrespective of $g$}.  For repulsive interactions ($g<1$), this regime corresponds to small voltages \cite{Kane92}.  Using the usual duality between large and small $g$ \cite{Weiss99}, this conclusion also holds for a QPC detector near perfect transmission, where the input signal modulates the strength of a weakly back-scattering impurity.  If the backscattering is weak enough that it can be treated to leading order (which requires large voltage for $g < 1$), one will again always be at the quantum limit. 

We next consider the $L \ra 0$ limit, but now consider regimes where higher order tunneling processes play a role.  Integrating out the phase $\phi$ in each term of our 
expansion of $\Z$, we obain a "Coloumb gas" description of $\Z$; such expansions have been used to great success in a variety of quantum impurity, and quantum dissipative problems \cite{Weiss99}.  We introduce auxiliary source fields in the action which couple to the Cooper-pair current and to the back-action force:
\begin{equation}
	\L_{src} = \sum_{\sigma=\pm} \sigma \left[
		\eta(t) \cdot I_{CPT}[\phi_{\sigma}(t)]
		+ \lambda(t) \cdot J[\phi_{\sigma}(t)]
		\right]
\end{equation}

To interpret the resulting expansion of $\Z$, it useful to make analogy to the Schmid model, a dissipative tight-binding (TB) model \cite{Schmid83}.  It describes a particle on a 
1-D tight-binding latticle which is coupled to a force produced by a bath of harmonic oscillators; this bath has a spectral density given by $A(\omega) = (8 e^2 / h) \omega Z_\phi(\omega)$.  In this mapping, 
$\tilde{E}_J = 2 E_J \cos \pi \frac{\Phi_x}{\Phi_0}$ corresponds to the tunnel matrix element of the TB model, and $2 e V_B$ to a constant external force applied to the particle.  The expression for $\Z$ (at $\eta, \lambda = 0$) may now be cast as a sum over tunnel events which take the particle from an initial density matrix localized at $x=0$ to one localized at $x=n$, where $n$ is arbitrary.  Each term in the expansion describes the amplitude of a process involving $2 M$ tunnel events occuring at times $t_1$ to $t_{2M}$.  Each event can move the particle either to the left or to the right, and can occur either on the forward or backwards Keldysh contour.  Each tunnel event is thus labelled by a charge $\sigma_j = \pm1$ which gives the direction of the tunnel event, and a charge $\xi_j$ which determines the contour on which the event occurs:  $\xi_j \sigma_j = 1$ (-1) for an event on the forward (backwards) contour.  Finally, 
because of decoherence from the bath,  each tunnel process results in a final density matrix state which is diagonal; we thus have the constraint $\sum \xi_j = 0$ for each term.  With these preliminaries, we have $\Z = \lim_{t \ra \infty} \sum_n P(n,t)$ with:
\begin{eqnarray}
	P(n,t) & = &  
		\sum_{M=n}^{\infty} 
			\left( \frac{i \tilde{E}_J}{\hbar} \right)^{2M} 
		\int_{-\infty}^{t}  dt_1
		...
		\int_{t_{2M-1}}^{t}  dt_{2M}
		\label{Eq:Pn}
		\\
	&&
		\sum_{ \vec \xi, \vec \sigma }
		\Bigg(
			\prod_{j=1}^{2M} 
				e^{i v \xi_j t_j}
			\cdot
			\left[ 1 + \sigma_j \xi_j \lambda(t_j) \right]
			e^{i \eta(t_j) \sigma_j}
		\Bigg)
		F[\vec{\sigma},\vec{\xi},\vec{\tau}] 
	\nonumber
\end{eqnarray}
Here, $n$ is the net displacement of a given process, and the factor $F$ describes interactions among the charges arising from integrating out the phase variable $\phi$; its precise form is given in Refs. \onlinecite{Ingold99, Weiss99, Baur04}.  The sums over charges in Eq. (\ref{Eq:Pn}) are restricted to those satisfying $\sum_j \xi_j = 0$ and $\sum_j \sigma_j = 2n$.

We see that the source field $\eta$ couples to the time derivative of the net displacement $n = \sum \sigma_j/2$ of the particle.  Thus, the Cooper-pair currrent $I_{CPT}$ corresponds to the velocity of the particle in the TB model.  This fact has been used previously to calculate the current noise in this system \cite{Saleur01}. We also see that the source $\lambda$ couples to $\sigma_j \chi_j$.  Thus, the back-action force corresponds to the time derivative of the ``quantum charge" $z = \sum \sigma_j \chi_j/2$; $z$ is the net number of forward contour minus backward contour tunnel events.  Note that for a given $M$ and $n$, $z$ may range from $-(M-|n|)$ to $M-|n|$.    

To make further progress, we follow Refs. \cite{Saleur01, Baur04}, and consider the Laplace transform representation of $\Z$.  As detailed in Refs. \cite{Saleur01, Baur04},  one may then discuss $\Z$ in terms of ``irrreducible clusters".  
Letting $s$ denote the Laplace transform variable, these are tunnel processes (i.e. a set of $\xi$ and $\sigma$ charges) which, in the $s \rightarrow 0$ limit, yield a finite constant contribution to $s [d \Z / dt](s)$.  Unlike Refs. \cite{Saleur01, Baur04}, we track both the classical displacement $n$ and the quantum charge $z$;  each irreducible process is characterized by a particular value of $n$ and $z$.  In the long time limit,  each such process is statistically independent, with its amplitude $\gamma_{n,z}$ acting as an independent rate;  one obtains Poissonian statistics for both $n$ and $z$.  At T=0 and finite bias, one finds that all rates with $n<0$ vanish.  One can also show $\gamma_{n,-z} = (\gamma_{n,z})^{*}$.  A straightforward calculation then yields:
\begin{eqnarray}
	\frac{ \langle I_{CPT}  \rangle } { 2e } & = & 
		\sum_{n=0}^{\infty}  n \left( \sum_z \gamma_{n,z} \right)
		= \sum_{n=0}^{\infty} n \cdot \Gamma_{cl}(n) 
			\label{Eq:AvgICG}\\
	\frac{ S_{I_{CPT}} } { 4e^2 }  & = & 
		\sum_{n=0}^{\infty} n^2 \left( \sum_z  \gamma_{n,z} \right)
		=  \sum_{n=0}^{\infty} n^2 \cdot \Gamma_{cl}(n) 
			\label{Eq:SICG}\\
	\frac{S_{J}} { e^2 B}  & = & 
		\sum_{z={-\infty}}^{\infty} z^2 \left( \sum_{n=0}^{\infty}\gamma_{n,z} \right)
		=  	 \sum_{z} z^2 \cdot \Gamma_{q}(z)
			\label{Eq:SJCG}
\end{eqnarray}
where $B = -\tan^2(\pi \Phi_x / \Phi_0)$.  We can interpret $\Gamma_{cl}(n) = \sum_z \gamma_{n,z}$ as a ``rate" associated with the incoherent tunneling of $n$ Cooper pairs; each such process contributes to the current independently.  Moreover, we see that back-action noise results from an uncertainty in how tunnel events are distributed between the two Keldysh contours (i.e. $z$ fluctuates).

Given that many independent processes contribute to the currrent, one's first guess is that the system will no longer be at the quantum limit (i.e. $\chi > 1$),
as generically, there is lost information associated with averaging over the different transport processes.  This would in be in complete analogy to the case of a non-interacting QPC with many transverse channels \cite{Pilgram02, Clerk03}.  To address this issue, we first make use of a remarkable result found by Saleur and Weiss:  the rates 
$\Gamma_{cl}(n) = \sum_z \gamma_{n,z}$ are {\it exactly} proportional to $E_J^{2n}$, and contain no higher-order 
terms \cite{Saleur01}.  Formally, this relation results from remarkable cancellations of terms arising in perturbation theory, and is related to the underlying integrability of the model.  It implies 
$\Gamma_{cl}(n) = \gamma_{n,0}$, and
that irreducible processes involving {\it both} backwards and forward tunneling events make no net contribution to charge transport. 

Using this above result, along with $\lambda = \frac{d \langle I_{CPT} \rangle }{ d \Phi} 
=  -
\tan \left( \pi \Phi_x / \Phi_0 \right)  
\frac{ \pi \tilde{E}_J}{\Phi_0} 
\frac{d \langle I_{CPT} \rangle } { d {\tilde{E}_J} }$, 
we see that a sufficient condition for having $\chi = 1$ is 
$\Gamma_{cl}(n) = - 2 \textrm{Re } \Gamma_{q}(z=n)$.  We have explicitly calculated the rates $\gamma_{n,z}$ to order $(E_J)^6$, which involves calculating 10 partial rates $\gamma_{n,z}$,
each of which has contributions from numerous charge sequences.  From this lengthy calculation, we find that a much stronger relation is satisfied:
\begin{eqnarray}
	\textrm{Re } \gamma_{n,z} & = & 0  \textrm{ if {\it both} n and z are nonzero} 
	\label{Eq:RateCond1} \\
	- 2 \textrm{ Re } \gamma_{n=0,z}  & = & \gamma_{n=z,0} 
	\propto  (E_J)^{2 z}
	\label{Eq:RateCond2}
\end{eqnarray}
Details will be presented elsewhere.  Note the duality between transport in the classical (i.e. $n$) and quantum (i.e. $z$) directions.  The only irreducible process yielding a classical displacement $n$ has precisely $2 n$ tunnel events all occurring in the same direction.  Similarly, the only irreducible process increasing the ``quantum charge" by $\Delta z$ has exactly $2 (\Delta z)$ tunnel events on the same Keldysh contour.  The cancellations which yield these results, and thus a quantum-limited back-action, require a purely Ohmic bath spectrum.  This is always the case for the interacting QPC.  For the dc SQUID, one needs the Josephson frequency $v$ to be much smaller than the cut-off frequency $1 / (R C)$ in order to reach the quantum limit at strong tunneling.  Note Eqs. (\ref{Eq:RateCond1})-(\ref{Eq:RateCond2}) {\it are not} a direct consequence of the result for $\Gamma_{cl}(n)$ found in Ref. \onlinecite{Saleur01}.  Also note that using duality \cite{Weiss99, Ingold99}, our conclusions also apply to the case where a perturbative expansion in $E_J$ does not converge, as one can formulate an alternate convergent expansion for $\Z$ which may be analyzed in an analogous manner.

The remarkable cancellations that lead to the result of Eqs. (\ref{Eq:RateCond1}-\ref{Eq:RateCond2})
 are, similar to the result 
$\Gamma_{cl}(n) \propto (E_J)^{2n}$, a result of the integrability of the model studied here.  Integrability may be used to exactly calculate the current and current noise in this system \cite{Fendley95}.  The solution is given in terms of the scattering of quasiparticles in a boundary sine-Gordon model.  One calculates an energy dependent quasiparticle transmission coefficient $\mathcal{T}(\ve)$, and then uses this to calculate the current and noise via a Landauer type approach.  Note that for a {\it non-interacting QPC}, reaching the quantum limit requires the QPC transmission $T(\ve)$ to satisfy \cite{Pilgram02, Clerk03}:
\begin{equation}
	\frac{d T(\ve)}{d \Delta} = C \cdot T(\ve) \left(1-T(\ve)\right)
\end{equation}
where $\Delta$ is the strength of the QPC back-scattering potential, and $C$ is a constant.  Remarkably, the energy-dependent transmission $\mathcal{T}(E)$ for quasiparticles in the boundary sine-Gordon model {\it satisfies
the exact same equation} (see Eq. 13 of Ref. \onlinecite{Fendley95}).  This suggests a deep connection between integrability and the quantum limit.  Calculating the back-action noise explicitly using integrability remains a challenging open problem, as the quasiparticle excitations of the boundary sine-Gordon theory are not eigenstates of the back-action operator.  

In conclusion, we have calculated the back-action noise and measurement efficiency for both the dc SQUID amplifier and interacting QPC point contact.  Using a perturbative approach, we have shown that it is still possible to reach the quantum limit in regimes where multiple higher-order tunneling processes play a role.

\bibliographystyle{apsrev}
\bibliography{references}



\end{document}